\documentclass[apj,twocolumn]{emulateapj}

\usepackage{graphicx}
\usepackage{lscape}
\usepackage{amssymb}
\usepackage{subfigure}

\def\Lsun{{L_\odot}}
\def\Msun{{M_\odot}}
\def\msun{{\Msun}}

\begin{document}


\title{Re-Assembling the Sagittarius Dwarf Galaxy}


\author{M. Niederste-Ostholt, V. Belokurov, N.W. Evans, J. Pe\~narrubia
}
\affil{Institute of Astronomy, University of Cambridge, Madingley Road, Cambridge, CB3 0HA, United Kingdom}
\email{mno,vasily,nwe,jorpega@ast.cam.ac.uk}

\begin{abstract}
  What is the mass of the progenitor of the Sagittarius (Sgr) dwarf
  galaxy?  Here, we reassemble the stellar debris using SDSS and 2MASS
  data to find the total luminosity and likely mass. We find that the
  luminosity is in the range $9.6 -13.2 \times 10^7 L_\odot$ or $M_V
  \sim -15.1 - 15.5 $, with $70\%$ of the light residing in the debris
  streams. The progenitor is somewhat fainter than the present-day
  Small Magellanic Cloud, and comparable in brightness to the M31
  dwarf spheroidals NGC 147 and NGC 185.  Using cosmologically
  motivated models, we estimate that the mass of Sgr's dark matter
  halo prior to tidal disruption was $\sim 10^{10} M_\odot$.
\end{abstract}


\label{firstpage}

\section{Introduction}

The Sagittarius (Sgr) dwarf galaxy is the closest Milky Way companion
($\sim 24$ kpc from the sun) and is currently being disrupted in the
Galaxy's tidal field. Both leading and trailing tidal streams have
been detected using many different tracer populations, including A
stars~\citep{Ya00}, RR Lyraes~\citep{Vi01,Wa09}, red clump
stars~\citep{Ma98,Ma99} and main sequence turn-off
stars~\citep{MD01,Ne02}. Perhaps the most spectacular views of the Sgr
stream are given by the M giants (Majewski et al. 2003). A portion of
the stream around the North Galactic Cap has been traced to much
fainter magnitudes using data from the Sloan Digital Sky Survey (SDSS,
see Belokurov et al. 2006a ``The Field of Streams'') revealing a
prominent bifurcation into the so-called streams A and B. There has
been considerable effort in modelling the disruption of Sgr since the
stream stars are useful tracers for the shape of the Milky Way's dark
halo potential.  Nonetheless, this work has yielded ambiguous results.
Depending on which datasets are fit, oblate~\citep{Jo05},
prolate~\citep{He04}, spherical~\citep{Fe06}, or even
triaxial~\citep{La09} halos are preferred.

One reason for this ambiguity is that the structure, size and nature
of the Sgr progenitor are very uncertain.  The original luminosity and
mass of Sgr is larger than its luminosity and mass today by
amounts depending on the mass density (including any dark matter) and
orbital history of the system. A neat illustration of this is provided
by the work of Jiang \& Binney (2000), who used N-body simulations to
show that the current configuration of the core of Sgr can be produced
by a very wide variety of initial conditions. At one extreme, the Sgr
dwarf might initially possess a mass in excess of $\sim 10^{11} \msun$
and fall into the Galaxy from a distance in excess of 200 kpc.  At the
other extreme, it might be only $\sim 10^9 \msun$ and
start off at distances similar to its present apocenter of 60 kpc.

In this paper, we analyse the number of stars and the luminosity of
the present-day Sgr core and debris. By re-assembling Sgr, we provide
new constraints on its parameters prior to disruption.  We describe
the data sets used for this study in \S 2 and analyse the
colour-magnitude distributions of the stream and core as well as the
luminosity density along the leading and trailing arms in \S 3. We
discuss the implications of our work for estimating the mass of
Sagittarius and conclude in \S 4.

\section{Data}

We trace the Sgr stream using main sequence, red giant and horizontal
branch stars from the SDSS (York et al. 2000), together with M giants
in the Two Micron All-Sky Survey (2MASS). SDSS is an imaging and
spectroscopic survey that covers one-quarter of the celestial sphere
including an area of $\sim8400$ deg$^2$ around the North Galactic Cap,
which contains a portion of the leading arm of the Sgr stream. There
are also some imaging stripes that sparsely sample low Galactic
latitude fields ($\sim3200$ deg$^2$) which contain parts of the
trailing arm.  The imaging data are collected in five passbands
($u,g,r,i,z$), for which there are model isochrones~\citep{Gi04,Do08}
and photometric
transformations~\footnote{http://www.sdss.org/dr4/algorithms/sdssUBVRITransform.html}
readily available. Here, we use positions and magnitudes of stars
extracted from Data Release 7~\citep{Ab09} and corrected for
extinction using the maps of \cite{Sc98}.  The bright limit in the
$u$-band is about 14 mag, whilst the faint limit is about 22. The
final SDSS sample contains $\sim 3,740,000$ stars along the
Sagittarius stream and $\sim 3,200,000$ stars in a comparison field in
the North Galactic Cap region as well as $\sim 294,000$ stars along the stream and
$\sim 255,000$ stars in the comparison fields in the southern
stripes. We additionally use the positions and $J,H,K$ magnitudes of
$\sim 1,450,000$ stars in the 2MASS All-Sky Point Source Catalog. This
is complete down to roughly $K=15$ mag.

Finally, we use a dataset from \cite{Bel06} to study the properties of
the Sgr core. It comprises of two fields ($\sim 299,000$ stars on and
$\sim 57,000$ stars off the Sgr core) in two passbands $B$ and $V$. The
sample goes approximately $3$ magnitudes below the main sequence
turn-off and includes stars up to the red clump, blue horizontal
branch and M giants.

\begin{figure*}
	\centering
	\includegraphics[width=\textwidth]{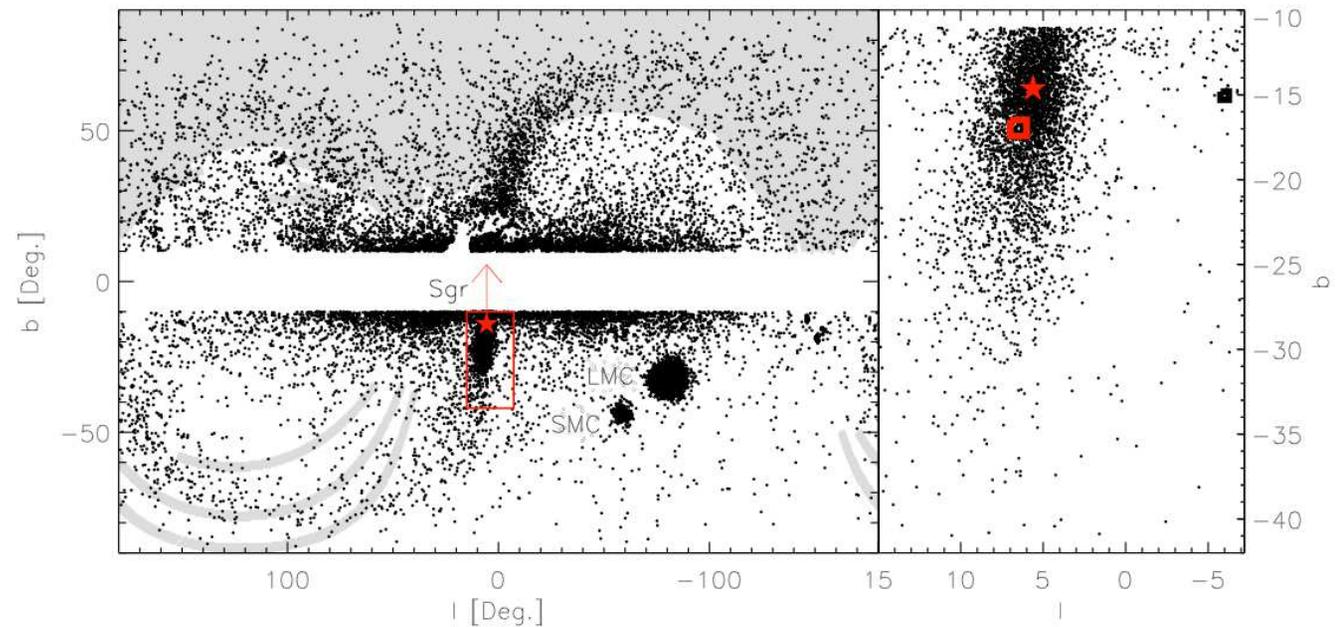}
	\caption{Left: M giants selected from 2MASS
          tracing the Sgr stream in Galactic coordinates. The box marks
          the central parts of the Sgr dwarf and the arrow indicates
          the direction of motion of the dwarf. The leading and
          trailing arms are visible at positive and negative Galactic
          latitude respectively. The gray-scale shows the SDSS
          footprint. Right: Zoom-in of the area outlined by the box,
          with the fields of \citet{Bel06} marked with squares. The
          photometric center of Sgr is marked with a star.}
	\label{fig:2mass_boxes}
\end{figure*}
\begin{figure*}
	\centering
	\includegraphics[width=\textwidth]{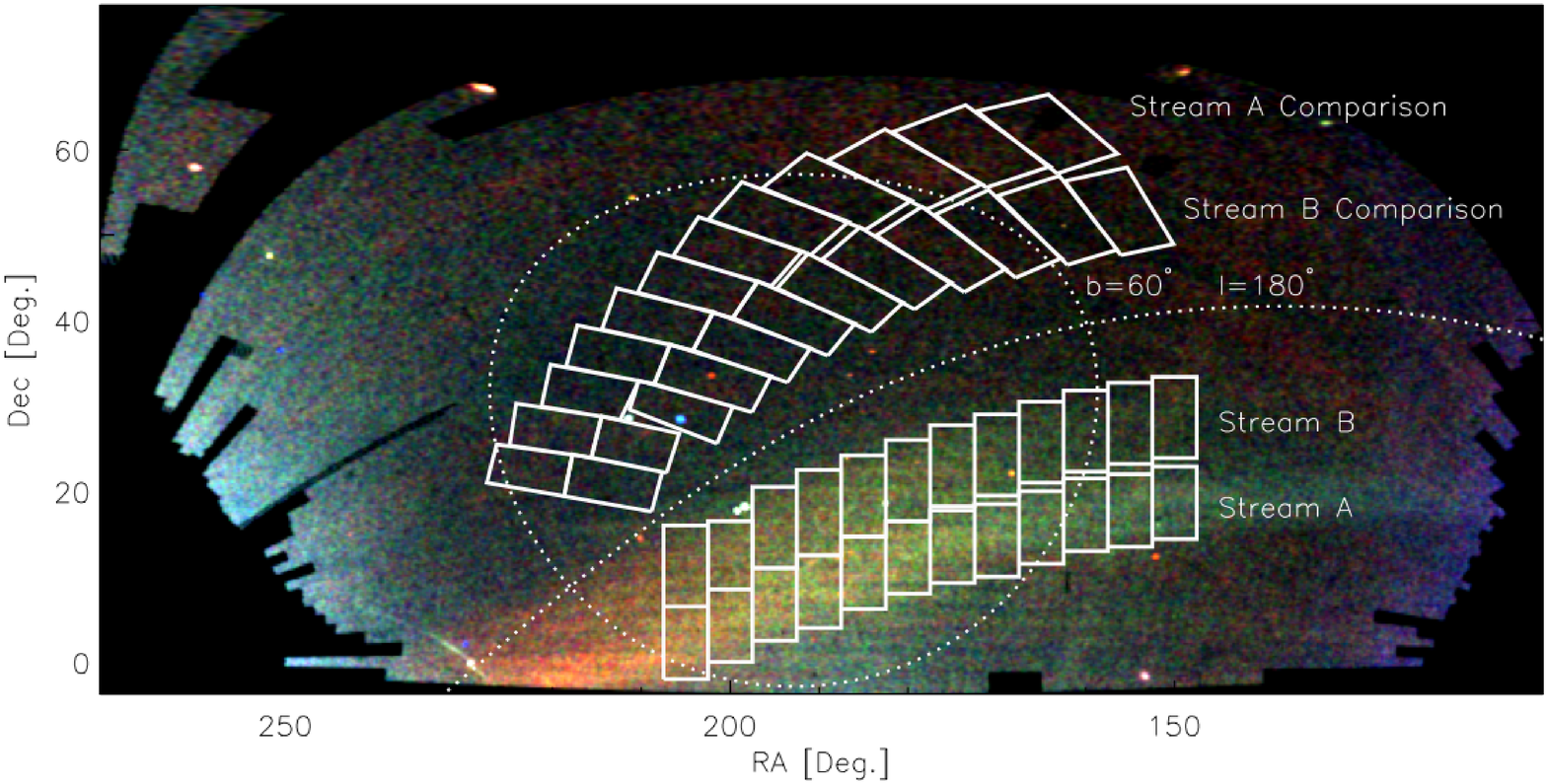}
	\caption{``Field of Streams'' \citep{Be06} with the Sgr debris
          bifurcated into a lower (A) and an higher (B) declination
          stream. Selection boxes along the stream (used in \S 3.3)
          are shown, together with comparison boxes at the same
          Galactic latitude but reversing the sign in longitude,
          (i.e., effectively reflecting the boxes in
          $\ell=180^\circ$). The stream boxes are $5^\circ$ wide in
          right ascension and are chosen to cover the entire width of
          the stream in declination.}
	\label{fig:fos_boxes}
\end{figure*}
\begin{figure*}
	\centering
	\includegraphics[width=\textwidth]{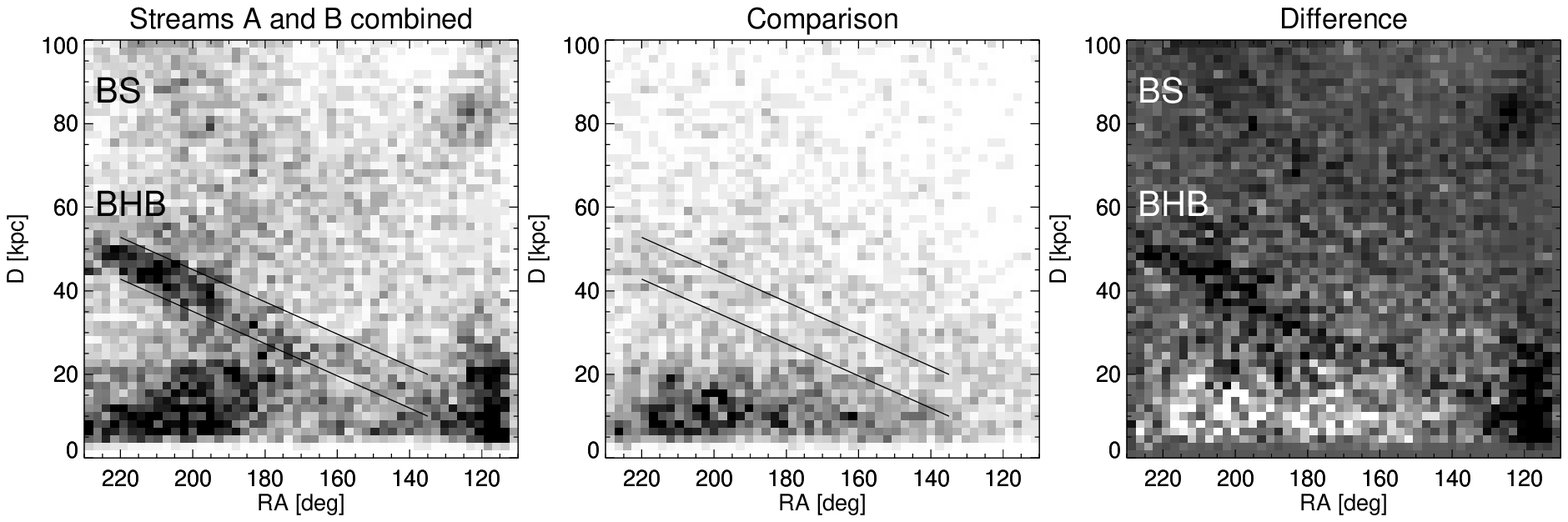}
	\includegraphics[width=\textwidth]{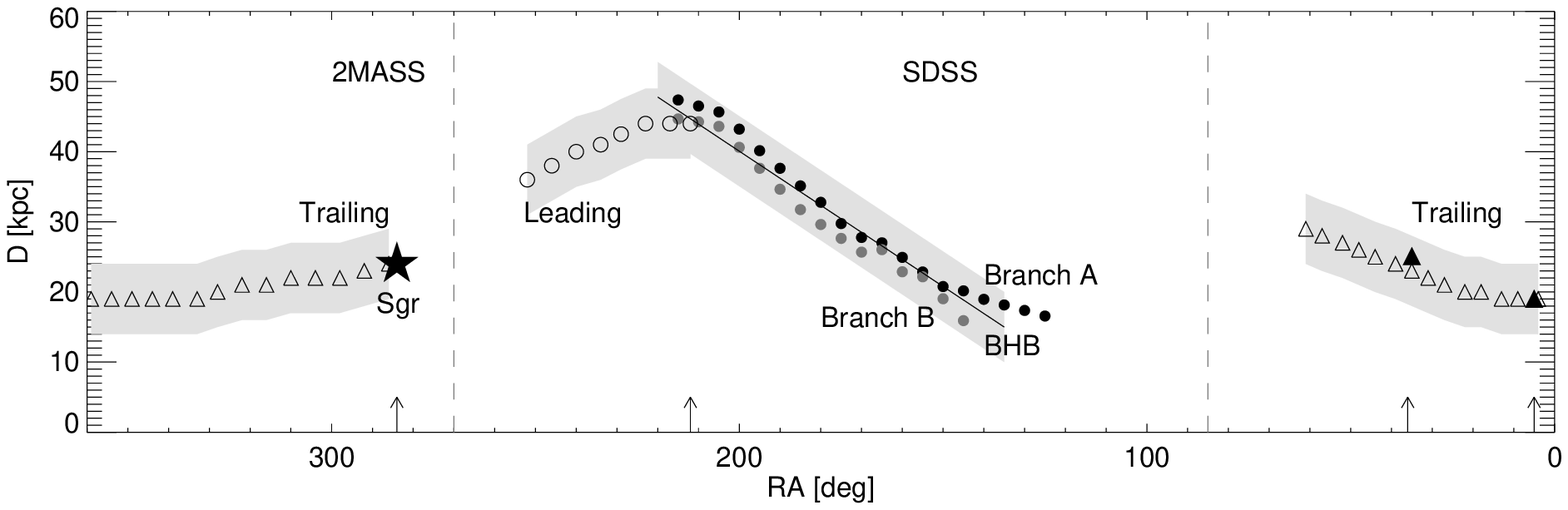}
	\caption{Upper: Density of SDSS BHBs in the plane of distance
          versus right ascension for the Sgr stream (left), comparison
          field (middle) and their difference (right). Note the
          prominent linear feature associated with streams A and B,
          which is shadowed by the blue stragglers some 2 magnitudes
          fainter.The solid lines bracket the distance gradient in
          Sgr's BHBs. Lower: Comparison of distances to the Sgr stream
          in 2MASS (open symbols) and SDSS (filled symbols, from
          Belokurov et al. 2006) datasets with the uncertainty in
          distance shown in gray. The leading arm is marked with
          circles, the trailing with triangles. Note the good match
          between distances derived from different indicators. Arrows
          mark locations of data-set overlaps.}
	\label{fig:distances}
\end{figure*}

\section{All-Sky Views of the Sgr Stream}

We begin by giving an overview of different observations of Sgr and
the terminology that we will employ throughout the paper. We refer to
the still bound portion of the dwarf as the core or remnant, which we
define as stars within 10 kpc of the center. These stars follow a King
profile (see e.g., Majewski et al. 2003).  The luminosity of Sgr is
usually quoted as the light integrated inside this radius, but it is
not obvious that all of these stars are still gravitationally
bound. The tidal debris from the disruption wraps around the Galaxy
(possibly multiple times) in what is usually referred to as the
leading and trailing tidal streams or arms. The portion of the debris
that is seen around the North Galactic Cap is bifurcated into stream A and stream
B. The working hypothesis for this paper is that both stream A and
stream B are part of the leading arm. Part of the leading arm is also
detected in the 2MASS survey. The trailing arm is in part covered by
the southern stripes of the SDSS and more fully covered in 2MASS.

The best panorama of the Sgr stream is given by M giants selected from
2MASS, as first realized by \citet{Ma03}. Their choice of cuts on
$J,H,K$ magnitudes picks out bright red stars at the typical distances
of Sgr.  For Figure~\ref{fig:2mass_boxes}, and in the following
analysis, we adopt the selection criteria introduced by Majewski et
al. (2003), namely
\begin{eqnarray}
\label{eq:thecuts}
J-K>0.85 \nonumber \\
J-H<0.561(J-K)+0.36 \nonumber \\
J-H>0.561(J-K)+0.22\\
|b| > 10^\circ \nonumber\\
E(B-V)<0.555\nonumber
\end{eqnarray}
Additionally, for the left panel of Figure~\ref{fig:2mass_boxes} only,
we remove some of the reddest stars with $J-K>1.25$ and limit the
magnitude (heliocentric distance) range to $10 <K<12$ ($7 < D < 70$
kpc) for the trailing arm and to $11 < K <13$ ($5 < D < 110$ kpc) for
the leading arm. This is in order to isolate the tidal debris of the
Sgr. Notice too that the Magellanic Clouds are very
prominent. Superficially, Sgr already looks to be intermediate
between the Large and Small Cloud in size and luminosity.

The right panel of Figure~\ref{fig:2mass_boxes} shows a zoom-in of the
Sgr core, made using the cuts of eqn~(\ref{eq:thecuts}), together with
an additional color-magnitude (CMD) box. This is derived by shifting
the ridgeline of Sgr M giants ($K=-8.650(J-K)+20.374$, Equation 5 of
Majewski et al. 2003) by $\pm3.5 \sigma$ or $\pm1.26$ mag in the color
range $0.95<J-K<1.1$. Using this selection gives a clear view of the
Sgr core, the photometric center of which is marked. Also shown are
the on-core and off-core fields of \citet{Bel06}.

The SDSS view of the Sgr stream is confined primarily to the North
Galactic Cap and hence is shown in equatorial coordinates in
Figure~\ref{fig:fos_boxes} to minimize distortion. The ``Field of
Streams''~\citep{Be06} shows the Sgr stream over a 100$^\circ$ arc,
with an unprecedented level of detail. The datasets provided by 2MASS
and SDSS are complementary -- the 2MASS survey picks up the trailing
arms with good clarity, but it loses the leading arm around the North
Galactic Cap.

\subsection{Distances}

There is information about the distances to debris at a number of
points along the stream~~\citep{Be06,Ne07,Wa09}. Heliocentric
distances to the trailing arm were originally computed by
\citet{Ma03}, who showed that there was little distance gradient along
this part of the stream. There has been more controversy about
distances to the leading arm, mainly because there are much steeper
distance gradients.

Given the crucial role that distances play, we begin by seeking
independent confirmation of earlier results. For the leading arm in
the Field of Streams, we select blue horizontal branch (BHB)
candidates from the SDSS in the on and off-stream fields using the
criteria given in Figure 10 of \citet{Ya00}. This gives us $\sim
18700$ candidates in the Sgr stream and $\sim 8500$ candidates in the
comparison field.  We note that these numbers do not include all
genuine BHBs, and suffer from some contamination from blue stragglers
(BS). The upper panels of Figure~\ref{fig:distances} show the density
of BHBs in the two fields, together with their difference.  The
distances to BHBs are calculated assuming an absolute magnitude of
$M_g = 0.7$, which is appropriate for our color range ($u-g > 0.9$,
see Table 2 of \cite{Si04}). Running diagonally across the plot is the
signal from the BHBs in the Sgr stream. It is accompanied by a fuzzy
shadow of BSs two magnitudes fainter. The Galactic halo foreground
appears to be slightly asymmetric, most likely due to the presence of
the Virgo Overdensity~\citep{Ju08} in parts of the stream
field. However, the subtraction works well and leaves the Sgr stream
clearly visible.  The stream signal is roughly linear in the plane of
distance versus right ascension and can be described by a straight
line with gradient 0.386.  The solid lines in the two leftmost panels
show the straight line fit offset by $\pm 5$ kpc. The lower panel
compares the BHB distances (straight line) to the leading arm with
those originally inferred by \citet{Be06} using the $i$ magnitude of
the subgiant branch in the ``Field of Streams''. It is reassuring that
there is excellent agreement between the two independently calculated
distances, especially as the BHB's luminosity is almost independent of
age and metallicity.

\begin{figure}
	\centering
	\includegraphics[width=0.5\textwidth]{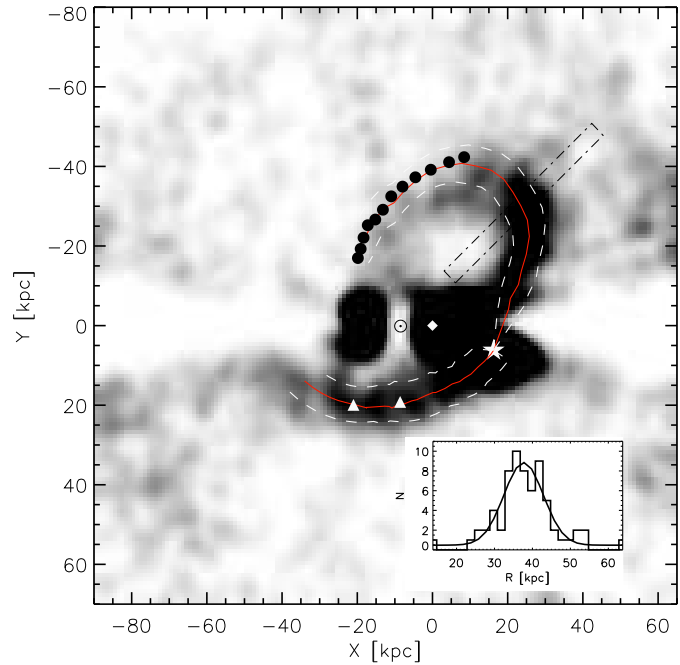}
        \caption{Density distribution of 2MASS M giants in the orbital
          plane of Sgr as defined in Majewski et al. (2003, see
          Figure 11).  The center of Sgr is marked with a star. The
          solid red line marks the ridge line, whilst the dashed lines
          indicate the $3 \sigma$ boundary of the tidal arms. The
          circles are the SDSS observations
          around the North Galactic Cap. The triangles mark the stream detections in
          the SDSS southern stripes. The inset shows the number of M
          giants as a function of heliocentric distance in the
          dot-dashed slice of the stream. A Gaussian fit to the
          distribution yields a width of $6$ kpc. The Galactic Center
          and the Sun positions are marked.} \label{fig:missing}
\end{figure}

Also shown are the distances to 2MASS M giants in both leading and
trailing arms. Using the cuts described above, we create a density
distribution of M giant stars in Sgr's orbital plane as shown in
Figure~\ref{fig:missing}.  Distances to pixels with highest density
along the stream are then computed with a typical uncertainty of 6
kpc. This is a combination of the uncertainty in the color-absolute
magnitude relation for M giants and the intrinsic width of the stream.
Finally, we also show two detections of the trailing debris in the
SDSS southern stripes. The distances are computed using both BHBs and
main-sequence turn-off stars. The overlaps of the SDSS and 2MASS
datasets will play an important role in what follows.

As a useful summary of the data, we list distances and right
ascensions in Table~\ref{tab:distances}. Recently, \citet{Ya09}
reviewed the detections of the Sgr stream in SDSS data, and their
results are in good agreement with ours.

\begin{figure*}
	\centering
	\includegraphics[width=\textwidth]{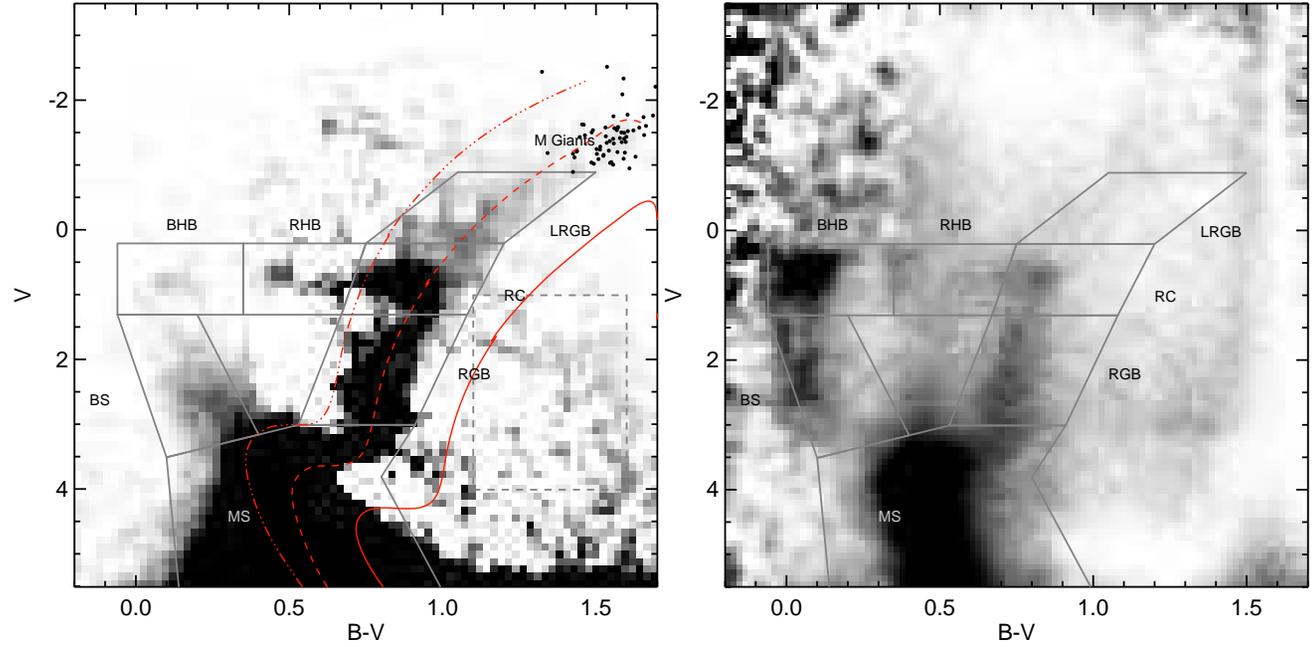}
        \caption{Comparison of the Hess diagrams of the Sgr core
          (left) and stream (right). Note that the populations in the
          core and the stream are very similar. Boxes are overlaid to
          identify characteristic populations used later in the paper,
          including main sequence (MS), red giant branch (RGB), red
          clump (RC), red horizontal branch (RHB), blue horizontal
          branch (BHB), blue stragglers (BS), and luminous red giant
          branch (LRGB). The black points in the left hand panel are M
          giant stars selected in the core using the cuts described
          above. The dashed box shown in the left panel is used to
          scale up the density in the control field. Overplotted in
          the left panel are Dartmouth isochrones ~\citep{Do08} of
          varying age and metallicity (solid $15$ Gyr, [Fe/H]$=0.0$;
          dashed $11$ Gyr, [Fe/H]=$-0.5$; dash-dotted $8$ Gyr,
          [Fe/H]$=-1.0$). We are confident that there are no turn-offs
          of older stellar populations in the Sagittarius core that
          were not detected in the Bellazzini et al. data.}
	\label{fig:hess_dif_sample}
\end{figure*}
\begin{figure}
	\centering
	\includegraphics[width=0.5\textwidth]{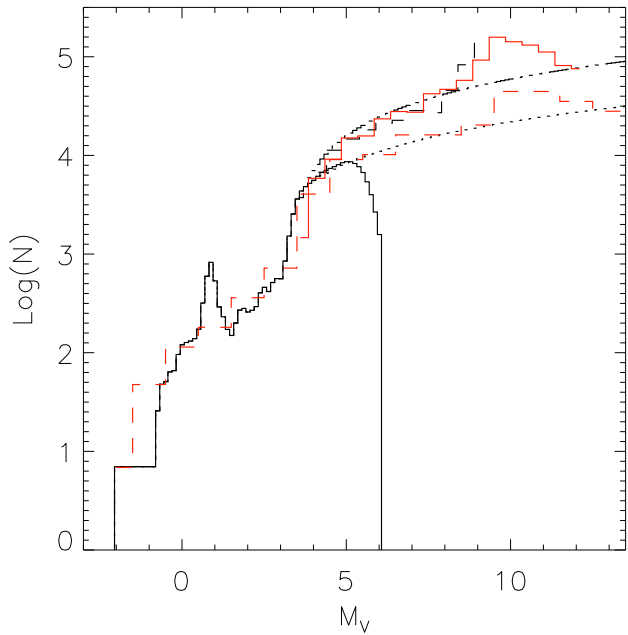}
        \caption{LF of the Sgr remnant (black), together with two
          possible extrapolations at the faint end (dashed and
          dotted). Also shown are the LFs of Ursa Minor (black
          dashed), the globular cluster M15 (red solid), together with
          a mean globular cluster LF (red dashed).}
	\label{fig:core_lf}
\end{figure}
\begin{deluxetable*}{lcccccccccccccccccccccc}
\tablecolumns{22}
\tablecaption{Heliocentric Distances to parts of the Sgr stream. SDSS and 2MASS observations are summarized in the top and bottom halves respectively.}
\tablehead{}
\startdata
RA - Stream A &215&210&205&200&195&190&185&180&175&170&165&160&155&150&145&140&135&130&125\\
Distance (kpc)&47&47&46&43&40&38&35&33&30&28&27&25&24&21&20&19&18&17&16\\ \\
RA - Stream B &215&210&205&200&195&190&185&180&175&170&165&160&155&150&145\\
Distance (kpc)&45&44&45&41&38&35&32&30&28&26&26&23&22&19&16\\ \\
RA - Trailing &5&35\\
Distance (kpc)&19&25\\ \\
\hline \\
RA - Leading &252 &246 &240 &234 &229 &223 &217 &212\\
Distance (kpc)&36&38&40&41&43&44&44&44\\ \\
RA - Trailing&286&292&298&...&316&322&328&333&...&18&22&27&31&35&39&44&48&52&57&61\\
Distance (kpc)&24&23&22&...&21&21&20&19&...&20&20&21&22&23&24&25&26&27&28&29\\ 
\enddata
\label{tab:distances}
\end{deluxetable*}

\begin{deluxetable*}{cccc}
\tablecolumns{4}
\tablecaption{Single exponential fits to the A (top half) and B (bottom half) 
  streams (see Figure~\ref{fig:boxcounts}).
  The left column gives the slope of the exponential together with its uncertainty. The middle column gives the luminosity per kpc of the fits at RA$=220^\circ$. The right column indicates the number of data points used in the exponential fit. Where applicable the values in brackets indicate the parameters derived using all data points.}
\tablehead{
\colhead{Sample}&
\colhead{Exponent ($\times 10^{-3}$)}&
\colhead{Luminosity  ($L_{\odot} \times10^4$)}&
\colhead{Data Points}
}
\startdata
Total & 7.0 $\pm$ 0.2 & 27.00 & 12\\
MS & 6.6 $\pm$ 0.3 & 15.92 & 12\\
BS & 4.5 $\pm$ 0.8 & 0.24 & 12\\
RGB & 6.4 $\pm$ 0.5 &2.01 & 12\\
RC & 7.7 (3.5)$\pm$ 1.0 (0.4) & 2.14 (1.66)& 6\\
RHB & 8.1 $\pm$ 0.6 &3.18 & 12\\
BHB & 6.2 $\pm$ 1.0 &0.33 & 12\\
LRGB & 8.7 (-2.7)$\pm$ 0.7 (0.5) &2.12 (1.1)& 4\\
\\
\hline
\\
Total & 4.5 $\pm$ 0.6 &15.01 & 12\\
MS & 4.9 $\pm$ 0.6 &9.20 & 12\\
BS & 2.7 $\pm$ 0.7 &0.18 & 12\\
RGB & 4.1 $\pm$ 0.7 &1.47 & 12\\
RC & 4.4 (0.5) $\pm$ 1.2 (0.4)& 1.2 (0.92) & 0\\
RHB & 5.1 $\pm$ 1.0 &1.47 & 12\\
BHB & 4.5 $\pm$ 1.2 &0.24 & 12\\
LRGB & 4.4 (-1.0) $\pm$ 1.2 (0.5) &1.2 (1.2) & 0
\enddata
\label{tab:powerlaws}
\end{deluxetable*}

\subsection{Sagittarius Core}

In order to derive the luminosity of Sgr, we need to disentangle its
stars from the Galactic fore and background. We start by re-visiting
the calculation of the luminosity of the remnant. The original value
of \citet{IGI} is based on data from Schmidt photographic plates. With
an estimate of the extent of the remnant, together with a CMD of a
$2.5^\circ \times 2.5^\circ$ central field, they calculated an
absolute magnitude of $M_V \sim -13$. This relies on the similarity of
the CMDs of Sgr on the one hand and Fornax and the SMC on the
other. \citet{Ma98} acquired data on 24 fields of size $0.25^\circ
\times 0.25^\circ$ across the face of Sgr and built a number
density profile of main sequence stars. Then, using a surface
brightness normalization from ~\citet{Ma95}, this can be integrated to
give a total luminosity of the remnant as $M_V = -13.3$. However, due
to the size of the field of view ($0.25^\circ \times 0.25^\circ$) in
~\citet{Ma95}, the luminosity function (LF) does not extend brighter
than the red clump and does not include BHBs and BSs. So, the shape of
the surface brightness profile is well-constrained, as confirmed by
\citet{Ma03}, but the overall normalization is more uncertain.

The data of \citet{Bel06} cover a larger area than those of
\citet{Ma95} at the same location, and hence give a clearer view of
all the stellar populations.  The core data are obtained in a
$1^{\circ} \times 1^{\circ}$ field approximately $2^{\circ}$ east of
the Sgr center located at $(l,b)\approx (6.5^{\circ},-16.5^{\circ})$
and along the major axis. The control field is $0.5^{\circ} \times
0.5^{\circ}$ at $(l,b)\approx (-6.0^{\circ} \times -14.5^{\circ})$.
For both fields, we fix the reddening at $E(B-V) = 0.11$, which is the
mean value over the core field, as judged from the maps of
\citet{Sc98}.  The left panel of Figure~\ref{fig:hess_dif_sample}
shows the CMD of the Sgr core. This is a Hess difference that has been
constructed by subtracting from the core Hess diagram the control
field Hess diagram. To scale the control field density, we minimize
the residual counts in the dashed box shown in
Figure~\ref{fig:hess_dif_sample}.  There are a number of readily
identifiable populations in the CMD, including the main sequence (MS),
red giant branch (RGB), red clump (RC), red horizontal branch (RHB),
blue horizontal branch (BHB), blue stragglers (BS) and the luminous
red giant branch (LRGB), together with M Giants selected in the same
manner as in the right-hand panel of Figure \ref{fig:2mass_boxes}. We
use the CMD boxes shown to build the LF of the core and calculate
luminosity density profiles along the stream.  It is reassuring that
our core CMD looks very similar to Figure 1 of \citet{Bel06}, which
has been constructed from the same data, but with a different
algorithm.

Figure~\ref{fig:core_lf} shows the observed LF of the Sgr remnant as
derived from integrating the light in the CMD boxes.  This drops
dramatically beyond $M_V \sim 5$, and a number of possible
extrapolations at the faint end are shown. However, most of the
luminosity is in stars brighter than the turn-off. \cite{Ma95}
computed the luminosity of the Sgr remnant by integrating the LF
between the red clump and the upper-main sequence to get a surface
brightness of 25.8 mag arcsec$^{-2}$. They account for the missing
flux in fainter stars by extrapolating their LF to $M_V =13.2$ to get
$0.5$ mag correction. They do not explicitly account for stars
brighter than the red clump or bluer than the main-sequence turn-off.
Repeating the calculation for our LF derived using the Bellazzini et
al. (2006) data, we get a similar surface brightness in stars between
the red clump and the upper main sequence, namely $25.7$ mag
arcsec$^{-2}$. However, we also find the BHBs, BSs, LRGBs and M giants
contribute an additional $0.3$ mag. Two possible extrapolations at the
faint end are shown, which contribute $0.5$ mag (dot-dashed) and
$0.15$ (dotted) respectively. This can be compared to LFs of a typical
globular cluster (red dashed, Cox 1999), the dSph Ursa Minor (black
dashed, Wyse et al. 2002) and one of the most luminous globular
clusters, M15 (red solid, Piotto et al. 1997). Taking our lead from
the dSph LF suggests adding $0.5$ mag for faint stars to arrive at a
surface brightness of $24.9$ mag arcsec$^{-2}$ (compared to a value of
$25.3$ given in Mateo et al. 1995).

\begin{figure*}
	\centering
	\includegraphics[width=\textwidth]{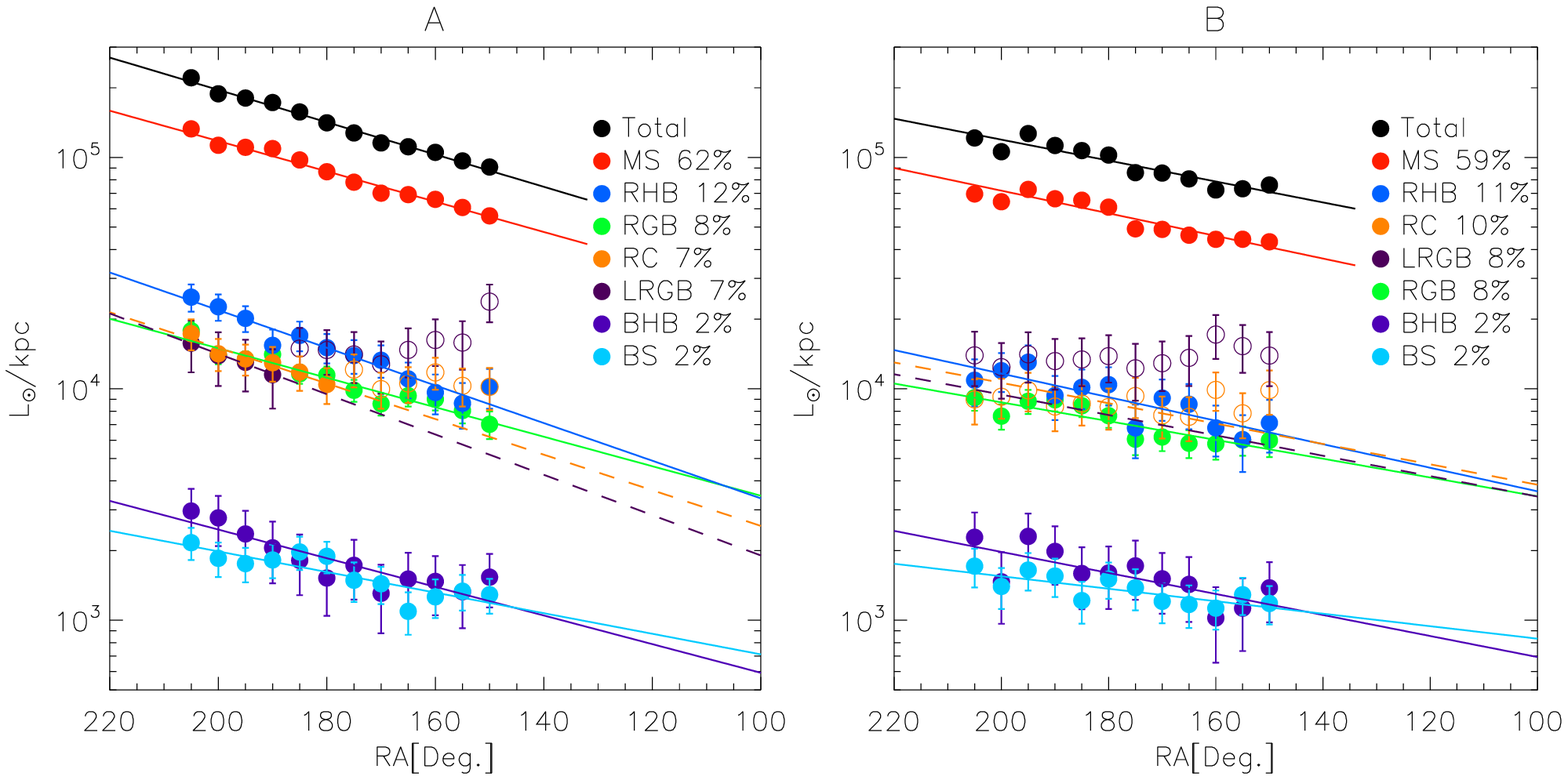}
	\caption{Left and right panels show the luminosity per unit
          length in a variety of stellar populations in streams A and
          B respectively. The luminosities for the RGB, RC, RHB, BHB,
          BS and LRGB stars are derived directly from the Hess
          difference diagrams. The numbers are normalized to $1$ kpc
          distance along the stream. The luminosities in the MS stars
          are corrected for missing flux using the derived LF. Error
          bars are determined assuming Poisson noise. A single
          exponential is fit to each population. However, the RC and
          LRGB populations show significant noise and hence for stream
          A only the first few data points follow an exponential
          fit. In stream B the noise dominates these two populations
          and the exponential shown is estimated without using any
          data points as described in the main text. The fits to these
          two populations are therefore shown as dashed lines. The
          data points used in determining the fits are shown as filled
          symbols and those excluded as open symbols. The behavior of
          streams A and B is very similar, suggesting that both A and
          B may be parts of the same leading arm. Finally, we list at
          the side of the figure the mean of the fractional
          contribution to the total luminosity across the stream for
          each population.}
	\label{fig:boxcounts}
\end{figure*}
\begin{figure*}
	\centering
	\includegraphics[width=0.475\textwidth]{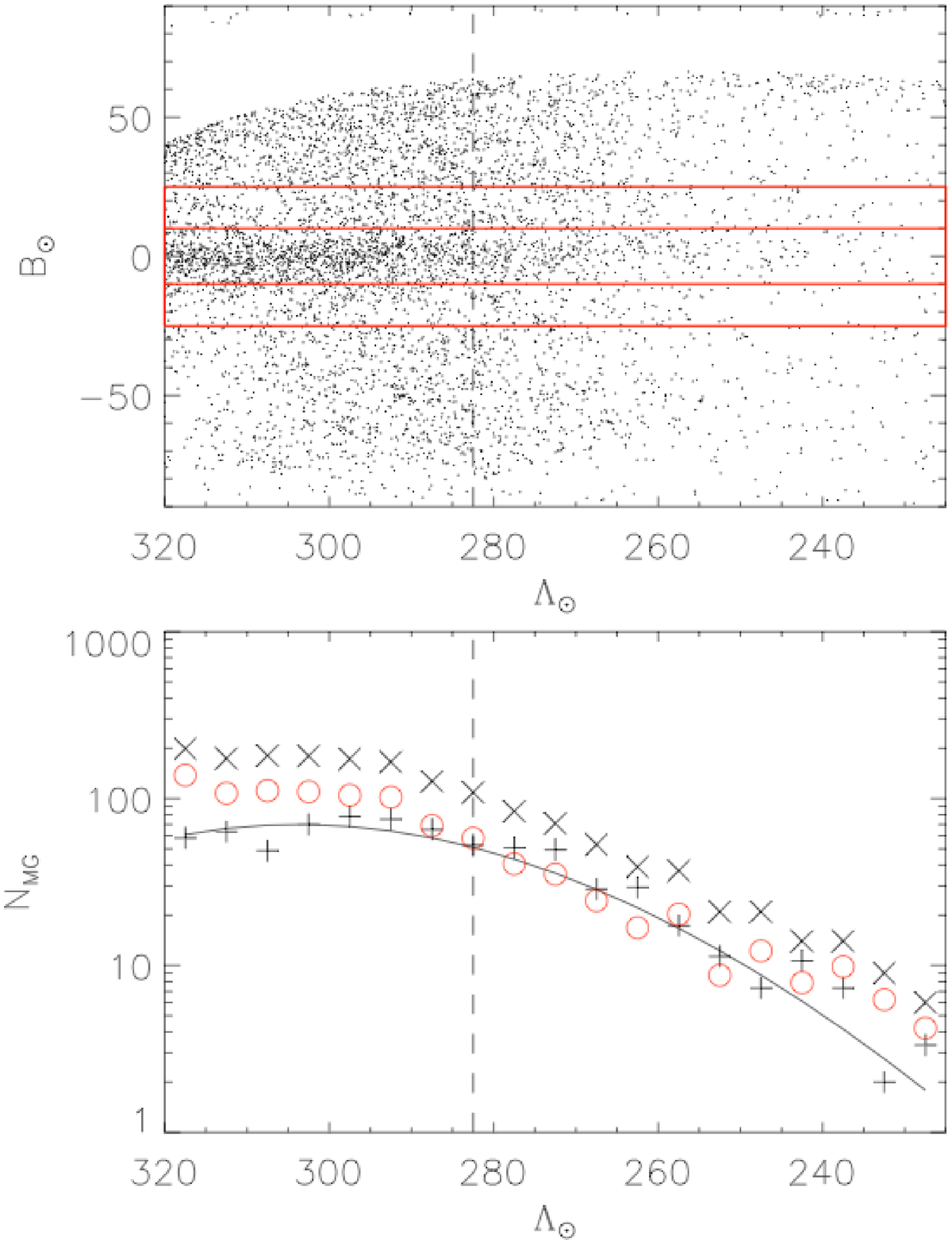}
	\includegraphics[width=0.475\textwidth]{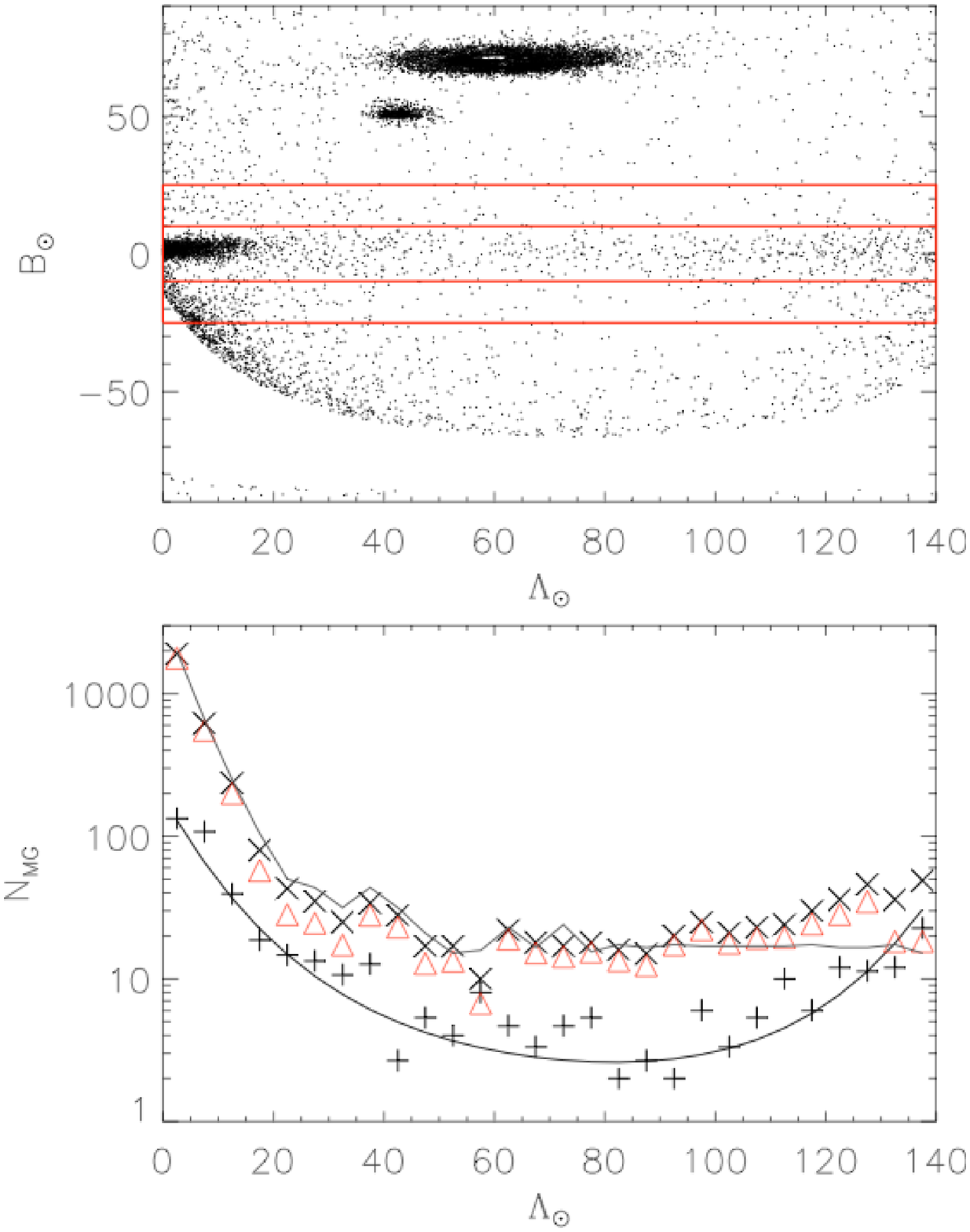}
	\caption{Top: Location of 2MASS M giants in the orbital plane
          defined by the Sgr orbit; see equation 9 of Majewski et
          al. (2003) for the definition of latitude and longitude
          ($B_\odot$,$\Lambda_\odot$). The red box encloses the Sgr
          leading (left) and trailing (right) streams, whilst the
          foreground box is chosen to have the same area. Bottom:
          Number counts ($N_{\rm MG}$) of M giants in the stream in
          bins of $5^\circ \times 20^\circ$ as a function of longitude
          in the orbital plane. Crosses (pluses) are counts in the
          central (foreground) boxes. The curves are estimates of the
          underlying smooth foreground. The red circles and triangles
          are the resulting profiles. The profile of the trailing arm
          derived by Majewski et al. (2003) is plotted in grey in the
          bottom right panel.}
	\label{fig:2MASSprofiles}
\end{figure*}

\subsection{The Density Profile of the Sagittarius Stream}

The CMD of streams A and B of the Sgr leading arm is already shown in
Figure 4 of \citet{Be06} and is known to look similar to that of the
core.  Here, we refine the calculation of the leading arm CMD in three
ways. First, we use the latest data release DR7 of the
SDSS~\citep{Ab09}. Second, we use a different comparison field,
defined by reflecting the boundaries of the stream across the
$\ell=180^{\circ}$ line (see Figure \ref{fig:fos_boxes}).  This
assumes that the Galactic field star population is symmetric across
$\ell = 180^\circ$ and that the control fields therefore probe a
similar field star population as is present around the Sgr stream.
Third, in order to combine portions of the stream at different
distances, we offset each star in the stream and control field using a
smooth mapping between its right ascension and distance (see Figure
\ref{fig:distances}). Rather than showing the difference of two Hess
diagrams, here we choose to show the ratio. The advantage of this is
that it accounts naturally for the fact that the depth of the CMD
varies along the stream. This procedure helps to enhance fainter
features like the horizontal branch and luminous part of the red giant
branch, as can be seen in the right panel of
Figure~\ref{fig:hess_dif_sample}.  This also explains the differences
in relative density between the two panels.  There are some small but
significant differences in the location of CMD features. For example,
the stream's CMD is slightly bluer than the core's. This could be due
to metallicity gradients, although the bulk of it is probably due to
slight underestimation of extinction in the core (which is of course
very difficult to measure in low latitude fields). Nonetheless, the
size of our CMD boxes means that regardless of the slight color
mismatch, Sgr member stars will still be selected.

We count the numbers of stream stars in the population boxes marked in
Figure~\ref{fig:hess_dif_sample}, and subtract the counts in the
comparison field. Given the distance to each field, this produces a
Hess difference in absolute magnitude versus color, $M_V$ versus
$B-V$.  Due to the substantial distance gradient along the stream, the
faintest population box -- main sequence stars -- is probed to
different limiting magnitudes. To correct this, we make use of the
fact that the stream's population is similar to that of the Sgr core,
at least below the subgiant branch. We extend each field's LF down to
$M_V=13.2$ by requiring that the slope of the stream's LF at $M_V =
3.4$ matches that of the core's extrapolated LF. This value is chosen
as it corresponds to the limiting SDSS magnitude $V \sim 22$ in the
most distant leading arm field.

The Hess difference is integrated over each CMD box to give the total
luminosity. Since each field has the same angular extent but is at a
different heliocentric distance, the luminosity is normalized by the
length of the segment in kpc. The two panels of
Figure~\ref{fig:boxcounts} show the luminosity profiles as a function
of right ascension for multiple populations in streams A and B,
together with exponential fits. The choice of fitting law is motivated
by the behavior of the total luminosity. We expect that all stellar
populations behave in a similar way, and this is largely confirmed by
our results. There are populations not so well described by an
exponential law, but these deviations may be caused by noise in the
data. Each of our fields probes at least some way down the
main-sequence, so the only population that is affected by our LF
correction is the main-sequence sample itself. However, as shown in
Figure \ref{fig:boxcounts}, the RGB, RHB, BHB and BSs all have slopes
that match the slope in the corrected main sequence sample, which
suggests that the observed trend is not due to the correction we
apply. Notice that the slopes of all the stellar populations in both
streams A and B are positive at a statistically significant
level. This implies that the corresponding densities of the
populations are all fading with decreasing right ascension, that is
increasing distance from the progenitor (as also noticed by Yanny et
al. 2009). This is consistent with both A and B being part of the same
leading stream.

Stream A is roughly a factor two brighter in luminosity than B. About
$60\%$ ($50\%$) of the signal is contributed by the main sequence
stars when the dot-dashed (dotted) extrapolation of Figure
\ref{fig:core_lf} is used. The total luminosities in the fields are on
average a factor of $1.3$ brighter when using the steeper
extrapolation. BHBs and BSs contribute little to the total luminosity,
but can be very cleanly selected and so give a good indication of the
underlying trend. By contrast, the counts corresponding to the RC and
LRGB stars are much noisier. This is because the foreground
contamination by the thick disc and halo is most severe, and varies
considerably in the bright part of the CMD.  In fact, the exponential
fall-off is only present in the first few bins of stream A,
corresponding to right ascensions between $180^\circ$ and
$205^\circ$. The dashed lines for stream A therefore show exponential
fits using only the data at higher right ascension (shown as filled
symbols). For the fainter and therefore even noisier stream B, even
this option is unavailable and so we assume a constant ratio of BHBs
to RCs and LRGBs in the two streams to derive the dashed lines in the
right-hand panel.  This assumes that streams A and B have similar
metallicities, which is supported by the work of Yanny et al. (2009).
Table \ref{tab:powerlaws} gives the slope of each exponential fit
together with its uncertainty, as well as the luminosity for each
component at a right ascension of $220^\circ$. We also indicate the
number of data points used in determining the exponential fit

To obtain as complete a profile of the stream as possible, we will
need to exploit the complementarity of the 2MASS and SDSS datasets.
2MASS provides a powerful probe of the trailing arm, for which
\cite{Ma03} have already deduced a profile. There are also parts of
the leading arm in 2MASS, for which we build the profile as
illustrated in Figure~\ref{fig:2MASSprofiles}. The upper panel shows
the locations of M giants, selected using the same method as the
zoom-in of Figure \ref{fig:2mass_boxes} after offsetting stars to the
distance of the core using Table~\ref{tab:distances}. The results are
shown in the heliocentric coordinate system aligned with the Sgr
orbital plane, as defined by equation 9 of \citet{Ma03}.  There are
two selection boxes in red to pick out the stream signal and the
foreground. The resulting profiles of the raw number counts in
$5^\circ$ bins, the foreground and the corrected number counts are
shown in the lower panel. Applying the same procedure to M giants in
the trailing arm, we recover a profile similar to that of
\citet{Ma03}, although rising in the outer parts.

\begin{figure}
	\centering
	\includegraphics[width=0.48\textwidth]{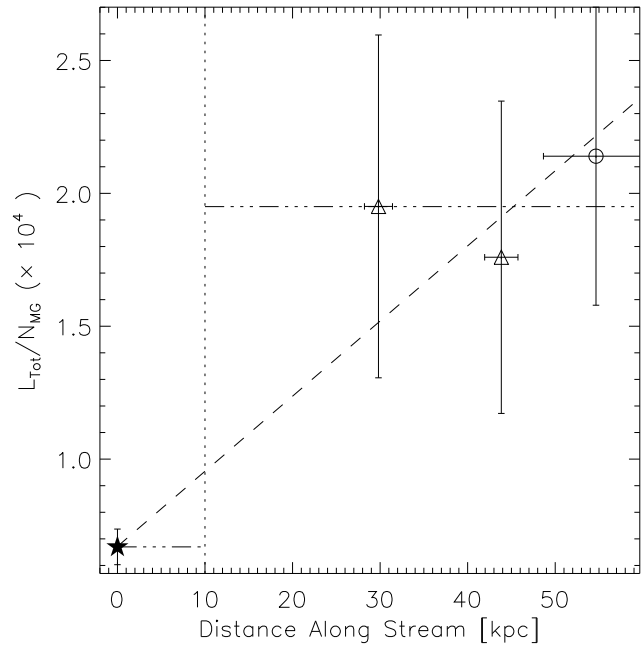}
        \caption{Conversion factor between total luminosity and
          numbers of M giants at different locations along the
          stream. The dash-dotted line shows the change in the
          conversion factor along the stream assuming that it is
          constant within the core, and constant along the stream. The
          dashed line shows the conversion factor assuming it varies
          linearly throughout the core and stream.}
	\label{fig:fields}
\end{figure}

Before we can begin reassembly, we need to do three more
things. First, we measure distance along the stream from the
progenitor to the observed fields using
Table~\ref{tab:distances}. Secondly, we combine streams A and B in the
SDSS dataset into a single profile of the leading arm. Thirdly, in
order to combine SDSS and 2MASS observations, we convert M giant
number counts into total luminosity densities along the stream. We
measure the relationship between the number of M giants ($N_{\rm MG}$)
and the total luminosity at the locations where the \citet{Bel06} and
SDSS observations overlap with the 2MASS coverage (see the lower panel
of Figure~\ref{fig:distances}). Here we can measure the total
luminosity using the Hess difference method and count 2MASS M giants.
The uncertainty in the ratio of the two is determined assuming Poisson
statistics. There are clearly substantial gradients between the
remnant and both the leading and trailing arms.  This is illustrated
in Figure~\ref{fig:fields}.  We use a straight line fit, which has a
gradient of $280 \pm 30 \Lsun$ kpc$^{-1}$ per M giant. Alternatively,
assuming a step function constant in the remnant and constant in the
stream does not affect the following analysis substantially.

\begin{figure}
	\centering
	\includegraphics[width=0.48\textwidth]{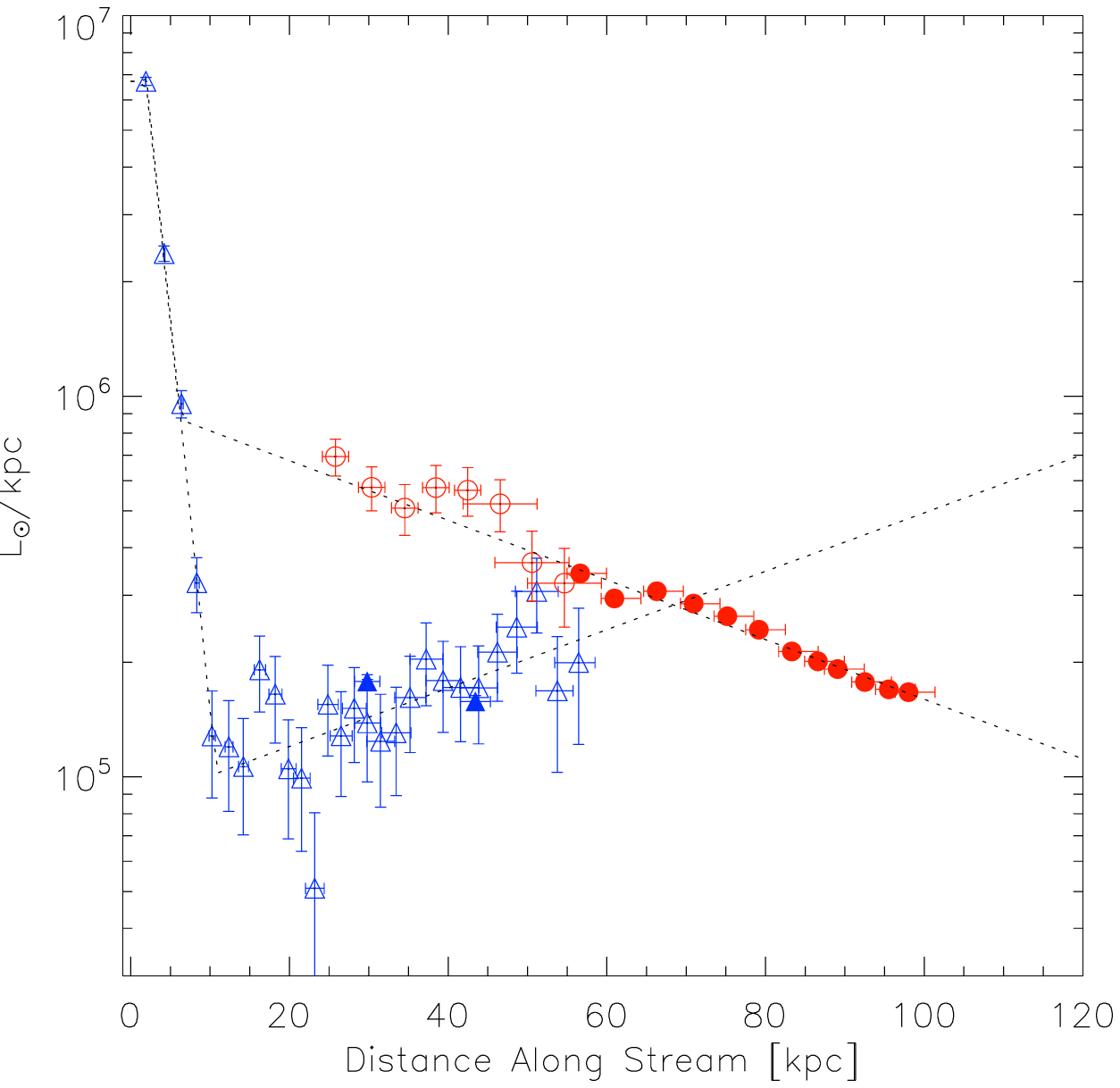}
        \caption{Comparison between the leading and trailing arms as
          seen in 2MASS and SDSS. We plot here the total luminosity
          per unit kpc along the stream versus the distance along the
          stream. For the 2MASS data, this is obtained from the number
          counts of M giants by the scaling described in the main
          text. The symbols have the same meaning as in the lower
          panel of Figure~\ref{fig:distances}. For the trailing
          stream, the dashed line shows the best broken exponential fit
          to the data. For the leading stream, the slope of the inner
          part is fixed which determines the break point.}
	\label{fig:luminosity_dist_profile}
\end{figure}

Figure~\ref{fig:luminosity_dist_profile} shows the resulting
luminosity profile as a function of distance along the stream for both
the leading (circles) and trailing (triangles) arms. Points derived
from 2MASS data are represented by open symbols, whereas those from
SDSS by filled symbols. The profiles of the leading and trailing arms
are different by a factor of $\sim 3$ at small distances from the
remnant, primarily as a consequence of the fact that the trailing arm
is much more extended.  Notice too the bump in the leading arm at
$\sim 40$ kpc distance along the stream. Such features are common in
simulations -- such as in Figure 1 of \citet{La05} -- as a consequence
of the pile-up of material stripped at disk-crossings. There is also
an evident break in the profile of the trailing arm at $\sim 10$ kpc
distance along the stream, similar to that seen by \citet{Ma03}. It is
natural to associate this break with the transition between the core
and the streams.  Even though the profiles have a complex structure,
it is useful to give a simple fit. For the trailing arm, we fit a
broken exponential and find indices $-0.207$ and $0.008$ either side of
a break radius at $9.6$ kpc. For the leading arm, the break radius
lies behind the disk and we only measure the profile after the break
as $-0.007$.

\subsection{The Luminosity of Sgr}

Now we begin the reassembly by calculating the total luminosity in the
Sgr remnant and tidal debris. We estimate the remnant's luminosity by
summing the observations of the trailing arm within the profile break
to get $2.0\pm0.3\times10^7 L_{\odot}$. This is doubled on the
assumption of the symmetric structure of the remnant, giving us a
total magnitude of $M_V = -14.2\pm0.1$, which is within 2 $\sigma$ of
the value of \citet{Ma98} of $M_V = -13.4\pm0.5$.

For the debris, we give two estimates. The first is a lower limit set
by the data alone, namely $2.6\pm0.1 \times 10^7 L_\odot$ for the
leading arm and $8.2\pm0.7 \times 10^6 L_\odot$ for the trailing.  The
combined value already contains more than half the core's luminosity,
and is well constrained by the deep SDSS coverage of the leading arm.

The second is an estimate of the luminosity contained within the
entire debris stream, not simply the portions we observe. This is
based on integrating the leading arm profile from $25$ kpc
(corresponding to the first 2MASS observation) to $300$ kpc, which is
triple the length to the last SDSS observation.  This corresponds to a
full wrap, as judged by the simulations~\citep[see
e.g.,][]{La05,Fe06}. This gives a luminosity of $3.5\pm0.1 \times 10^7
L_\odot$. This nominal error is so small because the fit is
constrained by the SDSS observations which have small uncertainties
due to the large number of stars in the SDSS sample. This also ensures
that potentially larger photometric errors on individual stars are
averaged out.  To account for the region between the core and $25$
kpc, we average two possible extreme cases.  If we assume that the
core is symmetric, then the leading arm profile needs to dip down
sharply to meet the trailing arm profile at $10$ kpc. This would
contribute $0.6\times10^7 L_\odot$ to the luminosity of the leading
arm.  Alternatively, since we do not actually observe any indication
of such a dip, the profile may continue rising until joining up with
the core's profile at approximately $5$ kpc.  This would contribute
$1.6\times10^7 L_\odot$. Therefore, for the missing inner portion of
the leading arm, we take as a compromise the average value
$1.1\pm0.5\times 10^7 L_\odot$ where the large uncertainty reflects
the range of possible values. The total leading arm luminosity (given
by adding the extrapolated exponential and the estimate for the region
between the core and the first observations)is then doubled to account
for the trailing arm. We do not integrate the profile of the trailing
arm since it is rising, which most likely reflects its approach to
appocenter. Integrating a rising profile is obviously very sensitive
to the limits of integration. At present it is difficult to make a
reasonable assumption about the behavior of the trailing arm far from
the core. This finally gives for the total debris luminosity
$9.2\pm0.7 \times 10^7 L_\odot$. This will be reduced to
$6.6\pm0.7\times10^7 L_\odot$ if the fainter LF in
Figure~\ref{fig:core_lf} is used.  It is evident that the formal error
from Poisson noise, typically of order of $10\%$, is much smaller than
uncertainties induced by extrapolations. To account for the missing
data, we were forced to extrapolate both the luminosity function and
parts of the density profiles.

The various luminosities used in the final answer are summarized in
Table \ref{tab:streamlum}. The total luminosity of Sgr progenitor is
$13.2\times10^7 L_\odot$ or $M_V \sim -15.5$ with the brighter
LF.  If we repeat the above calculation with the fainter LF, we arrive
at $9.6\times10^7 L_\odot$ or $M_V \sim -15.1$.  This implies
that the progenitor is fainter than the SMC, but comparable to the
brighter M31 dSphs like NGC 147 ($L=9.4\pm2.0\times10^7 L_\odot$,$M_V
= -15.1$) and NGC 185 ($L=13.5\pm2.0\times10^7 L_\odot$, $M_V =
-15.5$).

The systematic uncertainties clearly outweigh the Poisson noise but we
are confident that we have atleast found a firm lower limit
($9.6\times10^7 L_\odot$) to the luminosity. The sources of systematic
uncertainty include the interpolation of the inner slope of the
leading arm. This is over quite a modest range and unlikely to be
seriously in error. In addition considerable uncertainty derives from
the extrapolation of the LF as shown above. Perhaps the greatest
unknown is the behavior of the profiles at large distances. We have
assumed that the exponential behavior continues to 300 kpc, but there
may be processes that cause additional clumps and bumps.

\begin{deluxetable}{lccc}
  \tablecolumns{3} \tablecaption{Luminosity of various parts of the
    Sgr stream together with their associated Poisson
      uncertainties.}  \tablewidth{0.5\textwidth} \tablehead{{}&
    \multispan{2}{Luminosity (in $10^7 L_{\odot}$)} \\
    {Stream Portion} & {Faint LF} & {Bright LF}} \startdata
  Core (Bellazzini et al.) &$2\times1.5\pm0.3$ & $2\times2.0\pm0.3$ \\
  Leading Arm (SDSS) & $2.0\pm0.1$& $2.6\pm0.1$ \\
  Trailing Arm (2MASS) & $0.6\pm0.1$& $0.8\pm0.1$ \\
  Leading Arm (25 to 300 kpc) & $2.5\pm0.1$& $3.5\pm0.1$ \\
  Leading Arm (Core to 25 kpc) & $0.8\pm0.5$& $1.1\pm0.5$ \\
\enddata
\label{tab:streamlum}
\end{deluxetable}

\section{Discussion and Conclusions}

Using SDSS and 2MASS data, we have provided for the first time the
luminosity profiles of both the leading and trailing arms of the Sgr
stream.  Prior to this work, the only available profile was that of
the trailing arm derived from M giants by \citet{Ma03}. This was
essentially flat, and so cannot be extrapolated to derive useful
constraints on the mass lost during disruption. Here, our profiles for
both the leading and trailing arms have gradients. It is the gradient
in the leading arm that enables us to derive useful constraints.

The luminosity in the leading stream clearly declines with increasing
distance from the progenitor. Although the beginning of the tail is
absent because it lies behind the disk, the profile still shows mild
evidence for a bump. This is most likely a pile-up at apogalacticon.
The trailing stream can be traced all the way from the remnant to the
anti-Center, and shows a clear break around the tidal radius as we
move from the core to the tail. At first glance, this looks very
different to the leading arm profile. But in reality it may just be a
stretched-out version of the leading profile, as the trailing arm
extends to much greater Galactocentric distances. In this picture, the
rising part of the profile is interpreted as the approach to an
apogalacticon pile-up. We speculate that a similar but steeper rise
would be observed in the leading arm, making the core profile
symmetric, were it not obscured by the disk.

It is worth emphasizing that a firm lower limit $M_V \sim -15$ for the
Sgr progenitor luminosity is given by the data with minimal
assumptions. The most important is that the mass is equally
apportioned between leading and trailing in both the remnant and the
tails. Thanks to deep and wide SDSS coverage, the leading profile is
measured with excellent accuracy. It is the tightness of the
datapoints in the leading arm that is providing this firm lower limit.
For example, integrating the exponential profile out to $200$ or $300$
kpc makes little difference to the final answer.

Our results may have implications for the scenario proposed by
Fellhauer et al. (2006), who argued that stream A is material in the
leading arm that was stripped recently, whereas stream B is debris in
the trailing arm that was stripped long ago. Here, we have explicitly
assumed that streams A and B are both parts of the same leading arm.
This is supported by recent findings of \citet{Ya09}, who noted that
the stars in streams A and B have very similar metallicities and
velocities. Also, if stream B is old and trailing, then its density
(about one half of A) is larger than the young trailing arm (about a
third of A), which seems counter-intuitive.  So, although Fellhauer et
al's (2006) interpretation has not been ruled out, there are possible
causes for concern.

The luminosity profiles offer a new, and largely unexplored,
constraint on the disruption process. The fact that the leading arm
profile is not flat, but decreases, directly limits the stellar mass
lost in the disruption.  The relative numbers of counts in the leading
and trailing arms can be used to constrain the geometry of the
orbit. The stars move along the tail with a drift velocity determined
by the properties of the progenitor and the
orbit~\citep{De04}. Therefore, the time of onset of destruction of the
stellar component is in principle recoverable.

Our estimate of the progenitor luminosity is $9.6 - 13.2 \times 10^7
L_\odot$ or $M_V \sim -15.1 - -15.5$ with roughly $70\%$ of the light
in the tidal tails. The nearest look-alikes in the Local Group are
probably the M31 satellites NGC 147 and NGC 185.  These two dSphs
bracket properties of Sgr in terms of number of globular clusters,
metallicity, gas content and velocity dispersion~\citep[see
e.g. Tables 12.4 and 12.6 of ][]{Be00}. Their mass-to-light ratios are
comparatively modest at $\sim 5$~\citep{Ge09}. If we accept the
comparison, then this suggests that the mass within the luminous
radius of the Sgr progenitor was $\sim 10^9 \Msun$. This is right at
the lower mass end of the range of models considered by \citet{Ji00}
who investigated the range of progenitor properties which could
reproduce the structure of the core remnant observed today.

\begin{figure}
	\centering
	\includegraphics[width=0.4\textwidth]{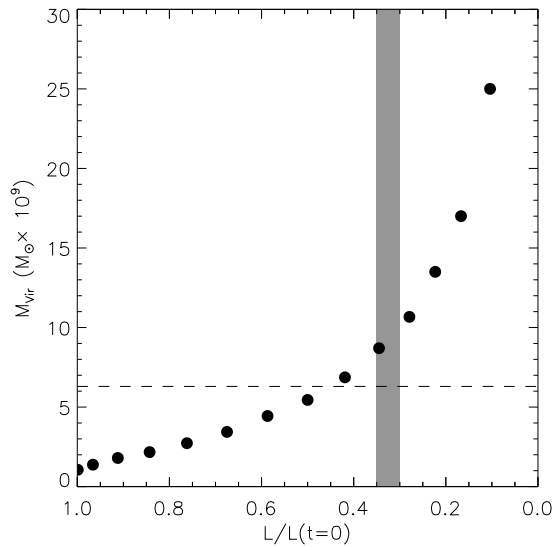}
        \caption{Virial mass of the NFW halo at $z=0$ as a
            function of luminosity remaining in the core. The shaded
            region shows the range of luminosities determined in this
            paper. If we are missing light from additional wraps or
            apocenter pile-ups the mass quickly exceeds that of the
            other classical dwarfs.The dashed line indicates the
            virial mass of Draco (the most massive classical dwarf
            according to Pe\~narrubia et al. $2008$a).}
	\label{fig:l_vs_mvir}
\end{figure}

We can use the estimate that $70\%$ of Sgr's luminosity now resides in
  the tidal debris to draw conclusions about the properties of the
  progenitor's dark matter halo. Pe\~narrubia et al. ($2008$,ab) have
  modelled the disruption of dwarf galaxies with baryons following a
  King profile embedded in a dark matter Navarro-Frenk-White (NFW)
  halo. The simulations suggest that the evolution of the parameters
  of the King profile is governed primarily by the amount of mass that
  has been stripped, rather than the details of the dwarf's orbit or
  the segregation of baryons and dark matter. Using as the present
  parameters of the bound core those determined by Majewski et
  al. ($2003$), we find that prior to disruption the King profile had
  a central velocity dispersion $\sigma_0= 23$ kms$^{-1}$, projected
  central stellar density $\Sigma_0= 5.7 \times 10^6 M_\odot$
  kpc$^{-2}$ and a core radius $R_{\rm C} = 1.5$ kpc. This implies
  that prior to mass stripping the central velocity dispersion,
  projected central stellar density and core radius were,
  respectively, a factor $2.0$, $5.3$, and $1.2$ higher than at
  present. Subsequently, we apply the method of Pe\~narrubia et
  al. ($2008$a), to estimate the properties of the NFW dark matter
  halo in which Sgr was embedded prior to mass stripping implied by
  this King profile. These authors solve the Jeans equations in order
  to find a set of $r_{\rm c,max}-v_{\rm c,max}$ parameters (the
  maximum circular velocity of the halo and the radius at which this
  is achieved which fully specify the NFW halo) that would allow a
  given velocity dispersion of the stellar component. This analyses
  yields a family of NFW models depending on the assumed spatial
  segregation of the stellar component within the dark matter halo. To
  break this degeneracy the authors appeal to the results of
  cosmological N-body simulations which show a strong correlation
  between an NFW halo's $v_{\rm c,max}$ and $r_{\rm c,max}$. Using the
  values for the King profile described above we derive a peak
  circular velocity $v_{\rm c,max}=35$ kms$^{-1}$ and scale radius
  $r_{\rm S}=3.5$ kpc ($r_{\rm c,max} \approx 2.17 r_{\rm S}$). The
  maximum circular velocity and scale radius are consistent with the
  values for the NFW haloes of other classical dSphs (see Figure $5$
  of Pe\~narrubia et al. $2008$a), but already towards the high mass
  end, with $M_{\rm vir} \approx 1.0 \times 10^{10} M_\odot$. The
  Draco dSph, found by Pe\~narrubia et al. to have the most massive
  dark matter halo, has $M_{\rm vir}\approx 6.3\times10^9
  M_{\odot}$. As shown by Pe\~narrubia et al. ($2006$), at this mass,
  dynamical friction can reduce the orbits apo- and pericenters by a
  factor of $>2$ over a Hubble time. We note that $9.6 - 13.2 \times
  10^7 L_\odot$ is likely a lower bound for the progenitor's
  luminosity since we have only considered one wrap of debris and have
  neglected the possibility of additional pile-ups at previous
  apocenters in the orbit. The amount of light contained in these will
  depend on the details of the mass loss process.  With additional
  light in the debris Sagittarius quickly becomes more massive than
  any of the other known Milky Way dwarfs (see Figure
  \ref{fig:l_vs_mvir}).

\acknowledgements 
MNO is funded by the Gates Cambridge Trust, the Isaac Newton
Studentship fund and the Science and Technology Facilities Council
(STFC), whilst VB acknowledges financial support from the Royal
Society. We thank Gerry Gilmore and Mike Irwin, as well as an
anonymous referee, for valuable comments.

Funding for the SDSS and SDSS-II has been provided by the Alfred P.
Sloan Foundation, the Participating Institutions, the National Science
Foundation, the U.S. Department of Energy, the National Aeronautics
and Space Administration, the Japanese Monbukagakusho, the Max Planck
Society, and the Higher Education Funding Council for England. The
SDSS Web Site is http://www.sdss.org/.

This publication makes use of data products from the Two Micron All
Sky Survey, which is a joint project of the University of
Massachusetts and the Infrared Processing and Analysis
Center/California Institute of Technology, funded by the National
Aeronautics and Space Administration and the National Science
Foundation.

\label{lastpage}

\end{document}